\documentclass[11pt,twoside]{article}

\usepackage{amsmath}
\usepackage{graphicx}
\usepackage{epsfig}
\usepackage[figuresright]{rotating}

\def\simge{\mathrel{%
   \rlap{\raise 0.511ex \hbox{$>$}}{\lower 0.511ex \hbox{$\sim$}}}}
\def\simle{\mathrel{
   \rlap{\raise 0.511ex \hbox{$<$}}{\lower 0.511ex \hbox{$\sim$}}}}

\def\ra{\!\rightarrow\!}
\def\dbar{\overline{D}{}^0}
\def\ycp{y^{}_{CP}}
\def\ycpKp{y^{K\pi}_{CP}}
\def\cp{$CP$}

\begin{document}

\title{
\begin{flushright}
{\normalsize UCHEP-22-03}
\end{flushright}
\vskip0.20in
{\bf {\boldmath Effect of $D^0$-$\dbar$ Mixing upon Cabibbo-favored \\
$D^0$ Decays }}
}

\author{\normalsize A.~J.~Schwartz \\
{\it Physics Department, University of Cincinnati, Cincinnati, Ohio 45221}}

\date{}

\maketitle

\begin{abstract}
The parameter $\ycp$ is used to characterize mixing in the $D^0$-$\dbar$ 
meson system. To determine $\ycp$, one measures the effective decay width of 
a $D^0$ or $\dbar$ decaying to a \cp\ eigenstate, relative to the effective width 
for a decay to a Cabibbo-favored, flavor-specific final state.
When using $\ycp$ to extract information about $D^0$-$\dbar$ mixing and \cp\ violation, 
the decay width of the Cabibbo-favored decay is usually assumed to equal $1/\tau$, 
the reciprocal of the $D^0$ lifetime. However, there is a small correction to this 
that should be taken into account when $\ycp$ is measured with sufficiently high 
precision. We calculate this correction in terms of charm mixing and \cp\ violation 
parameters $x$, $y$, $|q/p|$, and~$\phi$.
\end{abstract}

\section{\large Introduction}

An important parameter in the phenomenology of charm mixing is $\ycp$. 
It is defined as the difference from unity of the effective lifetime 
of $(D^0+\dbar)$ decays to a $CP=+1$ eigenstate, relative to the effective 
lifetime of $D^0$ or $\dbar$ Cabibbo-favored decay to a flavor eigenstate. 
Specifically~\cite{Liu,Bergmann},
\begin{equation}
\ycp\ \equiv\ \frac{\hat{\Gamma}[(D^0,\dbar)\ra K^+K^-]}{\hat{\Gamma}[D^0\ra K^-\pi^+]} \ -\ 1\,.
\label{eqn:ycpexp}
\end{equation}
The effective lifetimes are measured by fitting the corresponding decay time
distributions to exponential functions. When taking the ratio of effective
lifetimes in Eq.~(\ref{eqn:ycpexp}), many systematic uncertainties cancel 
out. Theoretically, for equal numbers of $D^0$ and $\dbar$ decays 
(as is the case, e.g., at an $e^+e^-$ collider), one finds~\cite{Bergmann}
\begin{equation}
\ycp\ =\ \frac{1}{2}\Bigl(\left|\frac{q}{p}\right| + \left|\frac{p}{q}\right|\Bigr) y\cos\phi
- \frac{1}{2}\Bigl(\left|\frac{q}{p}\right| - \left|\frac{p}{q}\right|\Bigr) x\sin\phi\,,
\label{eqn:ycpth}
\end{equation}
where $q$ and $p$ are complex coefficients relating the flavor eigenstates $D^0$ and 
$\dbar$ to the two mass eigenstates of the $D^0$-$\dbar$ system, and $\phi = {\rm Arg}(q/p)$. 
Equation~(\ref{eqn:ycpth}) shows that $\ycp$ is a combination of the mixing parameters 
$x=\Delta M/\Gamma$ and $y=\Delta \Gamma/(2\Gamma)$, where $\Delta M$ and $\Delta\Gamma$ are
the differences in masses and decay widths, respectively, between the mass eigenstates, and 
$\Gamma$ is their mean decay width. 

In calculating Eq.~(\ref{eqn:ycpth}), it is assumed that $\hat{\Gamma}(D^0\ra K^-\pi^+) = \Gamma$,
the mean decay width. In the past, this assumption was sufficiently accurate given the 
measured precision of $\hat{\Gamma}(D^0\ra K^+ K^-)$. However, a recent measurement of 
$\hat{\Gamma}(D^0\ra K^+ K^-)$ by the LHCb Collaboration~\cite{LHCb_ycp} has significantly 
greater precision than previous measurements~\cite{PDG_Dmixing} and thus 
requires more careful consideration of $\hat{\Gamma}(D^0\ra K^-\pi^+)$.
In particular, the effective decay width deviates from $\Gamma$ due to $D^0$ mesons 
that oscillate to $\dbar$ and subsequently decay to $K^-\pi^+$ via a doubly 
Cabibbo-suppressed amplitude. As the $D^0$-$\dbar$ mixing rate is very small, 
and the branching fraction for $\dbar\ra K^-\pi^+$ is only $1.4\times 10^{-4}$~\cite{PDG},
the effect of this process upon the decay time distribution is tiny and typically 
neglected. However, in Eq.~(\ref{eqn:ycpexp}), $\hat{\Gamma}(D^0\ra K^-\pi^+)$ is
compared to $\hat{\Gamma}(D^0\ra K^+ K^-)$, whose deviation from $\Gamma$ is also 
very small; thus the tiny effect in $\hat{\Gamma}(D^0\ra K^-\pi^+)$ can have an 
appreciable effect upon~$\ycp$. 

A more careful treatment of $\hat{\Gamma}(D^0\ra K^-\pi^+)$ adds a correction
term to Eq.~(\ref{eqn:ycpth}). Alternatively, one can define
$y_{CP}^{KK}\equiv [\hat{\Gamma}(D^0\ra K^+ K^-)/\Gamma] - 1$ and 
$\ycpKp\equiv [\hat{\Gamma}(D^0\ra K^- \pi^+)/\Gamma] - 1$, and then 
it is $y_{CP}^{KK}$ that equals the right-hand side of Eq.~(\ref{eqn:ycpth}).
The right-hand side of Eq.~(\ref{eqn:ycpexp}), which is what experiments 
measure, would essentially equal $y_{CP}^{KK} - \ycpKp$. To use
Eqs.~(\ref{eqn:ycpexp}) and (\ref{eqn:ycpth}) to extract 
information from $y_{CP}^{KK}$ about $x$, $y$, $|q/p|$, and $\phi$ 
(as done, e.g., by the Heavy Flavor Averaging Group~\cite{HFLAV}),
one must correct the measured value of $y_{CP}^{KK} - \ycpKp$ for~$\ycpKp$.

This paper presents a calculation of $\ycpKp$, or equivalently 
$\hat{\Gamma}(D^0\ra K^-\pi^+)$, the effective decay width of 
a Cabibbo-favored decay. This correction was first pointed out 
in Ref.~\cite{Pajero}. In that paper, $\ycpKp$ is calculated in terms 
of parameters $x^{}_{12}$, $y^{}_{12}$, $\phi^M_f$, and $\phi^\Gamma_f$, 
the magnitudes and phases of the off-diagonal elements of the $D^0$-$\dbar$ 
dispersive mass matrix and absorptive decay matrix~\cite{KaganSilvestrini}. 
In this paper, we calculate $\ycpKp$ in terms of the more common mixing 
and \cp\ violation parameters $x$, $y$, $|q/p|$, and $\phi$~\cite{PDG_Dmixing}.

\section{\large Calculation}

Starting from a pure $| D{^0}\rangle$ at $t=0$, the state found at a later time $t$ is
\begin{equation} 
|D^0(t)\rangle \ =\ g_+(t) |D^0\rangle\ +\ \left(\frac{q}{p}\right)\!g_-(t) |\dbar\rangle \,,
\end{equation}
where 
\begin{equation}
g_{\pm}(t) \ =\ \frac{1}{2} \left(e^{-i\omega^{}_1 t} \pm e^{-i\omega^{}_2 t}  \right)
\end{equation}
and 
\begin{equation}
\omega^{}_{1,2} \ \equiv\ m^{}_{1,2} - \frac{i}{2}\Gamma^{}_{1,2}
\end{equation}
are the eigenvalues of the two mass eigenstates $|D^{}_1\rangle$ and $|D^{}_2\rangle$.
The parameters $m^{}_{1,2}$ and $\Gamma^{}_{1,2}$ are the eigenvalues of the
Hermitian $2\times 2$ mass and decay matrices, respectively, and are real.
The decay amplitude to a final state $f$ is 
\begin{equation}
{\cal A}(D^0\ra f)\ =\ \langle f| H | D^0(t)\rangle\ =\ 
g_+(t)\,{\cal A}^{}_f \ +\ \left(\frac{q}{p}\right)\!g_-(t)\,\overline{\cal A}^{}_f\,,
\end{equation}
where ${\cal A}^{}_f \equiv \langle f| H | D^0 \rangle$ and 
$\overline{\cal A}^{}_f \equiv \langle f| H | \dbar\rangle$ are 
decay amplitudes for pure flavor eigenstates $D^0$ and $\dbar$.
The decay rate is thus
\begin{eqnarray}
r(t)\ =\ \Bigl|\langle f | H |D^0(t)\rangle \Bigr|^2
 & = &  \Bigl| g_+(t)\,{\cal A}^{}_f \ +\ \left(\frac{q}{p}\right)\!g_-(t)\,\overline{\cal A}^{}_f\Bigr|^2 \\
 & = & \bigl| {\cal A}^{}_f \bigr|^2
\Bigl| g_+(t)\ +\ \left(\frac{q}{p}\right)\!\frac{\overline{\cal A}^{}_f}{{\cal A}^{}_f}\,g_-(t)\Bigr|^2 \\
 & = & \bigl| {\cal A}^{}_f \bigr|^2
\Bigl| g_+(t)\ +\ \lambda\,g_-(t)\Bigr|^2 \\
 & = & \bigl| {\cal A}^{}_f \bigr|^2
\Bigl\{ \bigl| g_+(t)\bigr|^2 + \bigl|\lambda\,g_-(t)\bigr|^2
+ 2\,{\rm Re}\bigl[ \lambda\,g^*_+(t)\,g^{}_-(t)\bigr] \Bigr\}\,, \label{eqn:master1}
\end{eqnarray}
where (to reduce clutter) we have defined the parameter
 $\lambda\equiv (q/p)(\overline{\cal A}^{}_f/{\cal A}^{}_f)$.
Similarly, starting from a pure $| \dbar\rangle$ state at $t=0$, 
the decay rate to a final state $\bar{f}$ is
\begin{eqnarray}
\overline{r}(t)\ =\ \Bigl|\langle \bar{f} | H |\dbar(t)\rangle \Bigr|^2
 & = &  \Bigl| g_+(t)\,\overline{\cal A}^{}_{\bar{f}} \ +\ 
\left(\frac{p}{q}\right)\!g_-(t)\,{\cal A}^{}_{\bar{f}}\Bigr|^2 \\
 & = & \bigl| \overline{\cal A}^{}_{\bar{f}} \bigr|^2
\Bigl| g_+(t)\ +\ \left(\frac{p}{q}\right)\!
\frac{{\cal A}^{}_{\bar{f}}}{\overline{\cal A}^{}_{\bar{f}}}\,g_-(t)\Bigr|^2 \\
 & = & \bigl| \overline{\cal A}^{}_{\bar{f}} \bigr|^2
\Bigl| g_+(t)\ +\ \bar{\lambda}\,g_-(t)\Bigr|^2  \\
 & = & \bigl| \overline{\cal A}^{}_{\bar{f}} \bigr|^2
\Bigl\{ \bigl| g_+(t)\bigr|^2 + \bigl|\bar{\lambda}\,g_-(t)\bigr|^2
+ 2\,{\rm Re}\bigl[ \bar{\lambda}\,g^*_+(t)\,g^{}_-(t)\bigr] \Bigr\}\,, \label{eqn:master2}
\end{eqnarray}
where $\bar{\lambda}\equiv (p/q)({\cal A}^{}_{\bar{f}}/\overline{\cal A}^{}_{\bar{f}})$.

\vskip0.25in\noindent
We calculate the following:
\begin{eqnarray}
\left|g_+(t)\right|^2 & = & \frac{1}{4}\,\bigl|e^{-i\omega^{}_1 t} + e^{-i\omega^{}_2 t}  \bigr|^2 \nonumber\\
& = & \frac{1}{4}\,
\Bigl\{\bigl|e^{-i\omega^{}_1 t}\bigr|^2 + 
\bigl|e^{-i\omega^{}_2 t}\bigr|^2 + 2\,{\rm Re}\left[e^{i\omega^*_1 t}\,e^{-i\omega^{}_2 t}\right]\Bigr\} \nonumber \\
& = & \frac{1}{4}\,
\Bigl\{\bigl|e^{-i(m^{}_1-i\Gamma^{}_1/2)t}\bigr|^2 + 
\bigl|e^{-i(m^{}_2-i\Gamma^{}_2/2)t}\bigr|^2 + 
2\,{\rm Re}\left[e^{i(m^{}_1+i\Gamma^{}_1/2)t}\,e^{-i(m^{}_2-i\Gamma^{}_2/2)t}\right]\Bigr\} \nonumber \\
& = & \frac{1}{4}\,
\Bigl\{ e^{-\Gamma^{}_1 t} + e^{-\Gamma^{}_2 t} +  
2\,{\rm Re}\Bigl( e^{i \Delta m\,t}\,e^{-\overline{\Gamma}\,t}\Bigr)\Bigr\} \nonumber \\
& = & \frac{1}{4}\,
\Bigl\{ e^{-\overline{\Gamma}\,t} \Bigl[ e^{(-\Gamma^{}_1+\Gamma^{}_2) t/2} + e^{(-\Gamma^{}_2+\Gamma^{}_1) t/2}\Bigr] 
+ 2\,e^{-\overline{\Gamma}\,t} \cos(\Delta m\,t) \Bigr\} \nonumber \\
& = & \frac{e^{-\overline{\Gamma}\,t}}{2} 
\Bigl\{ \cosh\!\left(\frac{\Delta\Gamma}{2}\,t\right) + \cos(\Delta m\,t) \Bigr\}\,, \label{eqn:part1}
\end{eqnarray}
where $\Delta m\equiv m^{}_2 - m^{}_1$, $\Delta \Gamma\equiv \Gamma^{}_2 - \Gamma^{}_1$, 
and $\overline{\Gamma} \equiv (\Gamma^{}_1 +\Gamma^{}_2)/2$. Similarly,
\begin{eqnarray}
\left|g_-(t)\right|^2 & = & \frac{1}{4}\,\bigl|e^{-i\omega^{}_1 t} - e^{-i\omega^{}_2 t}  \bigr|^2 \nonumber\\
& = & \frac{1}{4}\,
\Bigl\{\bigl|e^{-i\omega^{}_1 t}\bigr|^2 + 
\bigl|e^{-i\omega^{}_2 t}\bigr|^2 - 2\,{\rm Re}\left[e^{i\omega^*_1 t}\,e^{-i\omega^{}_2 t}\right]\Bigr\} \nonumber \\
& = & \frac{1}{4}\,
\Bigl\{ e^{-\Gamma^{}_1 t} + e^{-\Gamma^{}_2 t} - 
2\,{\rm Re}\Bigl( e^{i \Delta m\,t}\,e^{-\overline{\Gamma}\,t}\Bigr)\Bigr\} \nonumber \\
& = & \frac{1}{4}\,
\Bigl\{ e^{-\overline{\Gamma}\,t} \Bigl[ e^{(-\Gamma^{}_1+\Gamma^{}_2) t/2} + e^{(-\Gamma^{}_2+\Gamma^{}_1) t/2}\Bigr] 
- 2\,e^{-\overline{\Gamma}\,t} \cos(\Delta m\,t) \Bigr\} \nonumber \\
& = & \frac{e^{-\overline{\Gamma}\,t}}{2} 
\Bigl\{ \cosh\!\left(\frac{\Delta\Gamma}{2}\,t\right) - \cos(\Delta m\,t) \Bigr\}\,. \label{eqn:part2}
\end{eqnarray}
Finally,
\begin{eqnarray*}
g^*_+(t)\,g_-(t) 
& = & \frac{1}{4}\,
\Bigl( e^{i\omega^*_1 t} + e^{i\omega^*_2 t}  \Bigr)
\Bigl( e^{-i\omega^{}_1 t} - e^{-i\omega^{}_2 t}  \Bigr) \nonumber \\
& = & \frac{1}{4}\,
\Bigl( e^{i(\omega^*_1-\omega^{}_1) t} - e^{i(\omega^*_2-\omega^{}_2) t} 
+ e^{i(\omega^*_2-\omega^{}_1) t} - e^{i(\omega^*_1-\omega^{}_2) t}  \Bigr) \nonumber \\
& = & \frac{1}{4}\,
\Bigl( e^{-\Gamma^{}_1\,t} - e^{-\Gamma^{}_2\,t}
+ e^{-i\Delta m\,t}\,e^{-\overline{\Gamma}\,t} - e^{i\Delta m\,t}\,e^{-\overline{\Gamma}\,t} \Bigr) \nonumber \\
& = & \frac{1}{4}\,e^{-\overline{\Gamma}\,t} 
\Bigl(
e^{(-\Gamma^{}_1 + \Gamma^{}_2)\,t/2} - e^{(-\Gamma^{}_2 + \Gamma^{}_1)\,t/2} 
+ e^{-i\Delta m\,t} - e^{i\Delta m\,t} \Bigr) \nonumber \\
& = & \frac{e^{-\overline{\Gamma}\,t}}{2} 
\Bigl\{ \sinh\left(\frac{\Delta\Gamma}{2}\,t\right) + i\sin\left(\Delta m\,t\right) \Bigr\} \,.
\end{eqnarray*}
Thus, 
\begin{eqnarray}
2\,{\rm Re}\bigl[ \lambda\,g^*_+(t)\,g^{}_-(t)\bigr] 
& = & e^{-\overline{\Gamma}\,t} 
\Bigl\{ {\rm Re}(\lambda )\sinh\left(\frac{\Delta\Gamma}{2}\,t\right) 
- {\rm Im}(\lambda)\sin\left(\Delta m\,t\right) \Bigr\}\,.
\label{eqn:part3}
\end{eqnarray}
Inserting Eqs.~(\ref{eqn:part1}), (\ref{eqn:part2}), and (\ref{eqn:part3}) into
Eq.~(\ref{eqn:master1}) gives
\begin{eqnarray}
r(t) & = & \frac{\bigl| {\cal A}^{}_f \bigr|^2}{2} e^{-\overline{\Gamma}\,t} 
\Bigl\{ 
(1+|\lambda|^2)\cosh\!\left(\frac{\Delta\Gamma}{2}\,t\right) + (1-|\lambda|^2)\cos(\Delta m\,t)
\nonumber \\
 & & \hskip1.5in 
+ 2\,{\rm Re}(\lambda )\sinh\left(\frac{\Delta\Gamma}{2}\,t\right) 
- 2\,{\rm Im}(\lambda)\sin\left(\Delta m\,t\right) \Bigr\}\,.
\label{eqn:master10}
\end{eqnarray}
Notating $\overline{\Gamma}$ as simply $\Gamma$ and defining 
mixing parameters $x\equiv\Delta m/\Gamma$ and 
$y\equiv\Delta \Gamma/(2\Gamma)$, Eq.~(\ref{eqn:master10}) becomes
\begin{eqnarray}
r(t) & = & \frac{\bigl| {\cal A}^{}_f \bigr|^2}{2} e^{-\Gamma t} 
\Bigl\{ 
(1+|\lambda|^2)\cosh(y\Gamma t) + (1-|\lambda|^2)\cos(x\Gamma t)  \nonumber \\
 & & \hskip2.0in
+ 2\,{\rm Re}(\lambda )\sinh(y\Gamma t) - 2\,{\rm Im}(\lambda)\sin(x\Gamma t) \Bigr\}\,. 
\label{eqn:master11}
\end{eqnarray}
The combination $\Gamma t = t/\tau^{}_{D^0}\equiv\tilde{t}$, 
the decay time in units of $D^0$ lifetime. For the range of $\tilde{t}$ measured 
by experiments, $x\tilde{t}\ll 1$, $y\tilde{t}\ll 1$, and we can use the 
following approximations:
$\cos(x\tilde{t})\approx 1-(x\tilde{t})^2/2$,
$\cosh(y\tilde{t})\approx 1 +(y\tilde{t})^2/2$,
$\sin(x\tilde{t})\approx x\tilde{t}$, and
$\sinh(y\tilde{t})\approx y\tilde{t}$.
With these approximations, Eq.~(\ref{eqn:master11}) becomes
\begin{eqnarray}
r(t) & = & 
\frac{\bigl| {\cal A}^{}_f \bigr|^2}{2} e^{-\Gamma t} 
\Bigl\{ 
(1+|\lambda|^2)\left(1+\frac{y^2}{2}\right) + (1-|\lambda|^2)\left(1-\frac{x^2}{2}\right)  
+ 2\,{\rm Re}(\lambda )y\tilde{t} - 2\,{\rm Im}(\lambda)x\tilde{t} \Bigr\}  \nonumber \\ \nonumber \\
& = & 
\bigl| {\cal A}^{}_f \bigr|^2 e^{-\Gamma t} 
\Bigl\{ 1 + \frac{y^2-x^2}{4} + |\lambda|^2\,\frac{x^2+y^2}{4} 
+ {\rm Re}(\lambda )y\tilde{t} - {\rm Im}(\lambda)x\tilde{t} \Bigr\} \nonumber \\ \nonumber \\
& = & 
\bigl| {\cal A}^{}_f \bigr|^2 e^{-\Gamma t} 
\Bigl\{ 1 + \frac{y^2-x^2}{4} + 
\left|\frac{q}{p}\right|^2 \left|\frac{\overline{\cal A}^{}_f}{{\cal A}^{}_f}\right|^2
\frac{x^2+y^2}{4} + \nonumber \\
 & & \hskip2.0in
\left|\frac{q}{p}\right|\,\left|\frac{\overline{\cal A}^{}_f}{{\cal A}^{}_f}\right|\,
\Bigl[ y\cos(\phi-\delta) - x\sin(\phi-\delta) \Bigr]\,\Gamma t \Bigr\} \,,  
\label{eqn:master13}
\end{eqnarray}
where $\phi \equiv {\rm Arg}(q/p)$ and 
$\delta\equiv {\rm Arg}({\cal A}^{}_f/\overline{\cal A}^{}_f)$.
While $\phi$ is a purely weak phase difference, $\delta$ is almost purely a 
strong phase difference: the weak phase difference between $D^0$ decay amplitudes 
is tiny due to charm decays proceeding almost exclusively via the 
first two flavor generations. 

\vskip0.25in\noindent
The various terms in Eq.~(\ref{eqn:master13}) have very different magnitudes.
The term $(y^2 - x^2)/4$ is quadratic in the small ($<1\%$) mixing parameters $x$ and $y$ 
and thus is negligible relative to the leading term. The term $(x^2 + y^2)/4$ is
also quadratic in mixing parameters; however, if the amplitude $\overline{\cal A}^{}_f$
is Cabibbo-favored and ${\cal A}^{}_f$ is doubly Cabibbo-suppressed, e.g., $f\!=\!K^+\pi^-$, 
then the term is greatly enhanced by the factor $|\overline{\cal A}^{}_f/{\cal A}^{}_f |^2$ 
and cannot be neglected. The last two terms, which are linear in mixing parameters, 
would be enhanced by the factor $|\overline{\cal A}^{}_f/{\cal A}^{}_f |$ and thus
should also be kept. These three terms yield the usual formula for the decay-time 
dependence of ``wrong-sign'' $D^0\ra K^+\pi^-$ decays; 
see Refs.~\cite{Kpi_babar,Kpi_belle,Kpi_CDF,Kpi_LHCb}.\footnote{For $f=K^+\pi^-$, our sign 
convention for $\delta$ is opposite that used for the strong phase $\delta$ in these papers.}
However, if ${\cal A}^{}_f$ is Cabibbo-favored and
$\overline{\cal A}^{}_f$ is doubly Cabibbo-suppressed, e.g., $f\!=\!K^-\pi^+$, 
then the $(x^2 + y^2)/4$ term can also be neglected. In this case,
Eq.~(\ref{eqn:master13}) becomes
\begin{eqnarray}
r^{}_{D^0\rightarrow K^-\pi^+}(t) & \approx & 
\bigl| {\cal A}^{}_f \bigr|^2 e^{-\Gamma t} 
\Bigl\{ 1 + 
\left|\frac{q}{p}\right|\,\left|\frac{\overline{\cal A}^{}_f}{{\cal A}^{}_f}\right|\,
\Bigl[ y\cos(\phi-\delta) - x\sin(\phi-\delta) \Bigr]\,\Gamma t \Bigr\} \nonumber \\
 & \approx & \bigl| {\cal A}^{}_f \bigr|^2 e^{-\Gamma t} \,e^{-y^{}_{K\pi}\,\Gamma t}\ =\ 
\bigl| {\cal A}^{}_f \bigr|^2 e^{-(1+y^{}_{K\pi})\,\Gamma t} \,,
\label{eqn:master14}
\end{eqnarray}
where 
\begin{eqnarray}
y^{}_{K\pi}\ =\ 
\left|\frac{q}{p}\right|\,\sqrt{R^{}_f}\,
%\left|\frac{\overline{\cal A}^{}_f}{{\cal A}^{}_f}\right|\,
\Bigl[ x\sin(\phi-\delta) - y\cos(\phi-\delta)\Bigr] 
\end{eqnarray}
with $R^{}_f\equiv |\overline{\cal A}^{}_f/{\cal A}^{}_f|^2$.
Equation~(\ref{eqn:master14}) implies that the decay-time distribution of Cabibbo-favored
$D^0\ra K^-\pi^+$ decays is essentially exponential, with a decay constant 
of $(1+y^{}_{K\pi})\times \Gamma$.

\vskip0.25in\noindent
For Cabibbo-favored $\dbar\ra K^+\pi^-$ decays, the decay rate $\bar{r}(t)$ is obtained
by comparing Eq.~(\ref{eqn:master2}) with Eq.~(\ref{eqn:master1}).  From this comparison,
we conclude that
\begin{eqnarray}
\bar{r}^{}_{\dbar\rightarrow K^+\pi^-}(t) 
 & \approx & \bigl| \overline{\cal A}^{}_{\bar{f}} \bigr|^2 e^{-(1+\bar{y}^{}_{K\pi})\,\Gamma t} \,,
\label{eqn:master15}
\end{eqnarray}
where 
\begin{eqnarray}
\bar{y}^{}_{K\pi}\ =\ 
\left|\frac{p}{q}\right|\,\sqrt{R^{}_{\bar{f}}}\,
\Bigl[ x\sin(-\phi-\delta) - y\cos(-\phi-\delta) \Bigr] 
\end{eqnarray}
with $R^{}_{\bar{f}} \equiv \left|{\cal A}^{}_{\bar{f}}/\overline{\cal A}^{}_{\bar{f}}\right|^2$.
If one selects a combined sample of Cabibbo-favored $D^0\ra K^-\pi^+$ and $\dbar\ra K^+\pi^-$ 
decays, with equal numbers of such decays, then the resulting decay-time distribution 
(assuming $|{\cal A}^{}_f|^2 = |\overline{\cal A}^{}_{\bar{f}}|^2$) will have a time 
dependence of
\begin{eqnarray}
e^{-\Gamma t}\left( e^{-y^{}_{K\pi}\,\Gamma t} + e^{-\bar{y}^{}_{K\pi}\,\Gamma t}\right) & = & 
2\,e^{-\Gamma t}\,e^{-(y^{}_{K\pi} + \bar{y}^{}_{K\pi})\,\Gamma t/2}\,
\cosh\left(\frac{\bar{y}^{}_{K\pi} - y^{}_{K\pi}}{2}\right)\Gamma t 
\nonumber \\
 & \approx & 2\,e^{-\Gamma t}\,e^{-(y^{}_{K\pi} + \bar{y}^{}_{K\pi})\,\Gamma t/2} \nonumber \\
 & \approx & 2\,e^{-\Gamma t}\,e^{-\ycpKp\,\Gamma t} \nonumber \\
 & \approx & 2\,e^{-(1+\ycpKp)\,\Gamma t}\,,
\label{eqn:master16}
\end{eqnarray}
where we have defined $\ycpKp\equiv (y^{}_{K\pi} + \bar{y}^{}_{K\pi})/2$.
This decay time distribution is also exponential, with a
decay constant differing from $\Gamma$ by the factor
\begin{eqnarray}
\ycpKp\ =\ \frac{y^{}_{K\pi} + \bar{y}^{}_{K\pi}}{2} & = & 
\frac{1}{2}\Bigl\{ \left|\frac{q}{p}\right|\,\sqrt{R^{}_f}
\Bigl[ x\sin(\phi-\delta) - y\cos(\phi-\delta) \Bigr] \nonumber \\
 & & \hskip0.50in
+\ \left|\frac{p}{q}\right|\,\sqrt{R^{}_{\bar{f}}}
\Bigl[ x\sin(-\phi-\delta) - y\cos(-\phi-\delta)\Bigr] \Bigr\} \nonumber \\
& = & 
\frac{1}{2}\left|\frac{q}{p}\right|\,\sqrt{R^{}_f}
\Bigl[ x\left(\sin\phi\cos\delta - \cos\phi\sin\delta\right)
- y\left(\cos\phi\cos\delta + \sin\phi\sin\delta\right) \Bigr] \nonumber \\
 & & \hskip0.50in
+\ \frac{1}{2}\left|\frac{p}{q}\right|\,\sqrt{R^{}_{\bar{f}}}
\Bigl[ -x\left(\sin\phi\cos\delta + \cos\phi\sin\delta\right)
- y\left(\cos\phi\cos\delta - \sin\phi\sin\delta\right) \Bigr] \nonumber \\
& = & 
-\frac{1}{2}\left(\left|\frac{q}{p}\right|\,\sqrt{R^{}_f} + 
\left|\frac{p}{q}\right|\,\sqrt{R^{}_{\bar{f}}}\right) 
\left( x\cos\phi\sin\delta + y\cos\phi\cos\delta\right)  \nonumber \\
 & & \hskip0.50in
+\ \frac{1}{2}\left(\left|\frac{q}{p}\right|\,\sqrt{R^{}_f} - 
\left|\frac{p}{q}\right|\,\sqrt{R^{}_{\bar{f}}}\right) 
\left( x\sin\phi\cos\delta - y\sin\phi\sin\delta\right) \,. \label{eqn:master3}
\end{eqnarray}
This expression is complicated. However, we can also apply this formula to
$f=K^+K^-$ or $\pi^+\pi^-$: 
for these self-conjugate final states, $\left|\overline{\cal A}^{}_f/{\cal A}^{}_f\right|=1$
and the $(x^2+y^2)/4$ term in Eq.~(\ref{eqn:master13}) can be neglected as done for
Cabibbo-favored decays. For $f = K^+ K^-$, $R^{}_f = R^{}_{\bar{f}} = 1$, 
$\delta =\pi$ (due to our phase convention $CP|D^0\rangle = -|\dbar\rangle$,
$CP|\dbar\rangle = -|D^0\rangle$), and Eq.~(\ref{eqn:master3}) simplifies to
\begin{eqnarray}
y^{KK}_{CP} & = & 
\frac{1}{2}\left(\left|\frac{q}{p}\right|+\left|\frac{p}{q}\right|\right) y\cos\phi - 
\frac{1}{2}\left(\left|\frac{q}{p}\right|-\left|\frac{p}{q}\right|\right) x\sin\phi\,.
\end{eqnarray}
This quantity is the well-known parameter~$y^{}_{CP}$~\cite{PDG_Dmixing,HFLAV},
i.e., we obtain Eq.~(\ref{eqn:ycpth}).

\vskip0.25in\noindent
As a final step, we utilize the fitted parameters of the Heavy Flavor Averaging Group (HFLAV)~\cite{HFLAV}
\begin{eqnarray}
R^{}_D & \equiv & \frac{|{\cal A}^{}_{\bar{f}}|^2 + |\overline{\cal A}^{}_f|^2}
         {|{\cal A}^{}_f|^2 + |\overline{\cal A}^{}_{\bar{f}}|^2 } \nonumber \\ \nonumber \\ 
A^{}_D & \equiv & \frac{|{\cal A}^{}_{\bar{f}}|^2 - |\overline{\cal A}^{}_f|^2}
                      {|{\cal A}^{}_{\bar{f}}|^2 + |\overline{\cal A}^{}_f|^2}\,\,,
\end{eqnarray}
and note that $R^{}_f = R^{}_D (1-A^{}_D)$ and $R^{}_{\bar{f}} = R^{}_D (1+A^{}_D)$.
Inserting these expressions into Eq.~(\ref{eqn:master3}) gives
\begin{eqnarray}
\ycpKp & = & 
-\frac{\sqrt{R^{}_D}}{2}\left(\left|\frac{q}{p}\right|\,\sqrt{1-A^{}_D} + 
\left|\frac{p}{q}\right|\,\sqrt{1+A^{}_D}\right) 
\left( x\cos\phi\sin\delta + y\cos\phi\cos\delta \right) \nonumber \\
 & & \hskip0.50in
+\ \frac{\sqrt{R^{}_D}}{2}\left(\left|\frac{q}{p}\right|\,\sqrt{1-A^{}_D} - 
\left|\frac{p}{q}\right|\,\sqrt{1+A^{}_D}\right) 
\left( x\sin\phi\cos\delta - y\sin\phi\sin\delta\right) \,. \nonumber \\
\label{eqn:master4}
\end{eqnarray}
To estimate how large $\ycpKp$ is, we insert values obtained
from the most recent HFLAV global fit~\cite{HFLAV_web}:
$x = 0.407$\%,
$y = 0.647$\%,
$R^{}_D = 0.344$\%,
$A^{}_D = -0.76$\%,
$|q/p| = 0.994$,
$\delta = 11.7^\circ$, and
$\phi = -2.6^\circ$.
%\begin{eqnarray*}
%x & = & 0.407\,\% \\
%y & = & 0.647\,\% \\
%R^{}_D & = & 0.344\,\%  \\
%A^{}_D & = & -0.76\,\% \\
%\left|\frac{q}{p}\right| & = & 0.994\\
%\delta & = & 11.7^\circ \\
%\phi & = & -2.6^\circ\,.
%\end{eqnarray*}
The result is $\ycpKp = -4.19\times 10^{-4}$, which has a magnitude $\ll\!1$ 
and thus is typically neglected in Eq.~(\ref{eqn:master16}).
However, to measure $y^{}_{CP}$, one compares the 
effective decay constant $\hat{\Gamma}$ of $D^0\ra K^+K^-$ or 
$D^0\ra\pi^+\pi^-$ decays with the effective decay constant of 
$D^0\ra K^-\pi^+$ or $\dbar\ra K^+\pi^-$ decays (or a combined sample), 
i.e., Eq.~(\ref{eqn:ycpexp}). The right-hand side of Eq.~(\ref{eqn:ycpexp}) equals
\begin{eqnarray}
\frac{\hat{\Gamma}^{}_{K^+K^-}}{\hat{\Gamma}^{}_{K\pi}} - 1 & = &
\frac{1+y^{KK}_{CP}}{1 + \ycpKp} - 1
\ \approx\ \bigl( 1 + y^{KK}_{CP}\bigr) \bigl( 1 - \ycpKp\bigr) -1 
\nonumber \\ \nonumber \\
 & \approx &  y^{KK}_{CP} - \ycpKp\,. \label{eqn:ycp_diff}
\end{eqnarray}
As the left-hand side of Eq.~(\ref{eqn:ycp_diff}) has recently been measured 
with an uncertainty of $2.9\times 10^{-4}$~\cite{LHCb_ycp}, it is necessary to 
account for $\ycpKp$ --\,which is larger than this uncertainty\,-- to 
determine $y^{KK}_{CP}$ (usually referred to as simply~$\ycp$).

\begin{center}
------------------
\end{center}

\noindent
The author thanks Joachim Brod and Marco Gersabeck for useful discussions 
and reviewing an earlier version of this manuscript. The author also thanks
Tommaso Pajero and Michael Morello for pointing out an inconsistency in notation.

\end{document}